\documentstyle[aps,preprint]{revtex}
\begin{document}
\draft 
\title {\bf Coulomb interaction in a quantum dot}
\author    {Lin-Wang Wang }
\address{National Renewable Energy Laboratory, Golden, CO 80401}
\maketitle
\begin{abstract}
 One approximation is made to describe a M+1 electron many-body 
 wavefunction by a M electron many-body wavefunction and a single electron
 wavefunction.
 Under this approximation, we have derived the Coulomb
 energy which relates the exciton energy $E_{exc}$
in a quantum dot with the quasiparticle band gap (defined as the difference
between the ionization energy and the electron affinity), and the Coulomb
energy which relates $E_{exc}$ with the single
particle eigen values. We found that these two Coulomb energies are different. 
We have compared our results with the formulae used in different groups, which 
are proposed either {\it ad hov}, or from the classical electrostatic point of view.  
We found important difference between our results and the classical 
formulae.
Finally, under the above approximation, 
we provided an effective single particle Hamiltonian, which
gives the quasiparticle band structure in a bulk system. 

\end{abstract}
\vskip 0.5in


\narrowtext

\centerline {\bf I. Introduction}

   The screened Coulomb interaction in a bulk semiconductor between an excited electron 
and a hole has been studied theoretically 30 years ago \cite{Abe,Sham} from many body point of views. 
As a result of such theoretical analysis, the screening of the Coulomb interaction 
can be expressed using the dielectric response function of the bulk system. 
Recently, nanometer scale finite system has attracted a lot of research interests
\cite{MRS}.
The confined excited electrons and holes within the finite systems have enhanced 
Coulomb interactions comparing to the bulk material \cite{Ashoori}. 
However, there are different opinions about how to screen 
the Coulomb interactions in various cases. While the screening of the
Coulomb interaction is clear when the electron is outside the nanostructure, it becomes
clouded from a classical point of view when it enters the nanostructure. One problem is
that the classical dielectric screening effects can already been partially represented by
the single particle eigen energies. Thus combining the classical electrostatic model with 
the single particle Hamiltonian does not always provide a clear picture free from ambiguities.
The more rigorous approach is to derive the single particle effective Hamiltonian
and the screened Coulomb interaction
from the many body Hamiltonian.
However, the Feyman diagram technique used in the bulk study
thirty years ago can not be directly applied to the finite system without 
complicated modifications of the Green's functions according to the new quantum dot boundary 
conditions. In this article, we will present a derivation of the screened electron-hole
Coulomb interaction in a nanosystem started from a many body Hamiltonian. Instead
of basing on the Green's functions, our derivation is based on many particle wavefunctions.
We will restrict ourselves to the
cases that the excited electrons and holes are strongly confined by a 0 dimension
nanostructure (quantum dot). That means, the correlation between the excited electron and the hole
can be ignored \cite{Yoffe,Japan} 

 One of the earlier works on 
the quantum dot Coulomb interaction and its screening effects is given by L.E. Brus \cite{Brus1}. 
Using a classical model of the dielectric screening and a single particle Hamiltonian, Brus derived
the change of electron affinity of a quantum dot (spherical, with radius R) 
relative to the bulk system: 

$$ E_N-E_{N+1}= E_{aff}^{bulk} - \epsilon_e - {1\over 2 R} \bar P'.    \eqno(1) $$
Here $E_M$ is the total energy of the quantum dot with M electrons. $E_N$ is the neutral quantum 
dot. Thus $E_N-E_{N+1}$ is the electron affinity of the quantum dot. $E_{aff}^{bulk}$
is the electron affinity of the bulk. $\epsilon_e$ is a single electron confinement energy in 
conduction band, which equals $\pi^2\over 2 m^*_e R^2$ under an effective mass model of electron 
effective mass $m^*_e$. ${1\over 2 R} \bar P'$ is an electrostatic energy representing the 
interaction between the electron and its image charge produced by the quantum dot medium with dielectric
constant $\epsilon$. Similarly, for ionization energy, he has

$$ E_{N-1}-E_N=E_{ion}^{bulk} + \epsilon_h + {1\over 2 R} \bar P'.   \eqno(2) $$
Here $\epsilon_h$ is the single hole confinement energy, which equals $\pi^2\over 2 m^*_h R^2$
under a single band effective mass model of hole effective mass $m^*_h$. The last term in Eq(2) is
the same as the last term in Eq(1). 
Then, using the conventional definition, the quasiparticle band gap $E^{qp}_g$ equals
the difference between the ionization energy the the electron affinity:

$$ E^{qp}_g \equiv E_{N+1}+E_{N-1}-2 E_N= E_g^{bulk}+\epsilon_e+\epsilon_h + {1\over R} \bar P'. \eqno(3) $$
Here, we have used the fact that the bulk band gap $E_g^{bulk}$ equals the difference between
bulk ionization energy $E_{ion}^{bulk}$ and the bulk electron affinity $E_{aff}^{bulk}$.
In this paper, we discuss only the cases where the quantum dot is surrounded by vacuum. Then 
if the dielectric constant of the quantum dot $\epsilon$ is much larger than 1, we have
(following Ref.\cite{Allan}):

$$ {1\over R} \bar P' = (1-{1\over \epsilon}) {1\over R} + {0.94 \over \epsilon R}
\big ( {{\epsilon-1}\over{\epsilon+1}} \big ).    \eqno(4) $$
In a second paper \cite{Brus2}, Brus presented the result for the exciton energy $E_{exc}$ in a quantum dot (which equals 
the photon energy needed to excite an electron 
from the valence band to the conduction band), 

$$ E_{exc}= E_g^{bulk}+\epsilon_e+\epsilon_h - {1.8\over \epsilon R} + {0.94 \over \epsilon R}
\big ( {{\epsilon-1}\over{\epsilon+1}} \big ),       \eqno(5)  $$  
where $- {1.8\over \epsilon R}$ represents the direct Coulomb integral between the electron charge
and hole charge, screened by $\epsilon$, using the effective mass wavefunctions.
The last term in Eq(5) is said to be the time
averaged instantaneous
dielectric energy while the classical particle moves around inside the quantum dot. 
Comparing Eq(3) to Eq(5), with the help of Eq(4), we have

$$E_{exc} = E^{qp}_g - {1.8\over \epsilon R} - (1-{1\over \epsilon}) {1\over R}.  \eqno(6) $$
Note that, classically, the last term in Eq(6) represents the Coulomb interaction energy between 
a spherical charge density inside the quantum dot with its
induced surface charge. The total of the last two terms in Eq(6) represents the electrostatic
interaction between the confined bare electron charge density with 
a screened hole charge (i.e, the total charge density of the $E_{N-1}$ system).

However, the above classical picture is not universally accepted. 
For example, in a recent paper
of Ogut, etal\cite{Ogut}, the connection between $E^{qp}_g$ and $E_{exc}$ is not given 
by Eq(6), instead a simple Coulomb interaction $- {1.8\over \epsilon R}$ is used (in the
effective mass limit):

$$E_{exc} = E^{qp}_g - {1.8\over \epsilon R}.  \eqno(7) $$
The difference between Eq(7) and Eq(6) is very large (up to 2 eV for $R\sim 7 \AA$), 
because there is an unscreened
Coulomb interaction $-1/R$ in the last term of Eq(6). Using Eq(7), Ogut etal obtained
an exciton energy $E_{exc}$ much larger than the one obtained from the single particle
empirical pseudopotential method (EPM) \cite{EPM} (which is yield via Eq(5) but without its last 
term, and with $\epsilon_e$, $\epsilon_h$ calculated via single particle EPM Hamiltonian). 
Based on this difference, they concluded that the empirical pseudopotential calculation
is wrong because a size dependence of the GW selfenergy $\sum$ is ignored. However, if
Eq(6) instead of Eq(7) is used, one finds that the result from Ref.\cite{Ogut} is almost the same as
the EPM result. Then their conclusion will be false and EPM calculation will be all right. 

To settle the dispute between Eq(7) and Eq(6), and to test the other classical formulae
[e.g, Eq(5)], it will be very useful to derive the above relationships from many body
Hamiltonians. This will be provided in the rest of this paper. We found out that Eq(6)
is correct instead of Eq(7), and Eq(5) is correct after deleting its last term.

\vskip 0.5in
\centerline {\bf II. The basic Formalism}

   In this section, we will derive our basic formula to be used in later sections. 
 This formula is based on an approximation that the many-body wavefunction 
 $\Phi_{M+1}$ of a M+1 electron system can be separated into one single particle
 wavefunction $\psi$ [i.e, $\psi_v$ (top of valence band state) 
 or $\psi_c$ (bottom of conduction band state)] and the rest of the M particle 
 wavefunction $\Phi_M$.  More specifically, we have:

$$\Phi_{M+1}(x_1,x_2,...,x_M,x_{M+1})\equiv|\Phi_M\psi>\equiv{1\over\sqrt{M+1}} \big[
\Phi_M(x_1,x_2,...,x_M) \psi(x_{M+1})$$
$$~~~~~ -\Phi_M(x_{M+1},x_2,...,x_M) \psi(x_1)
... -\Phi_M(x_1,x_2,...,x_{M+1})\psi(x_M) \big].  \eqno(8)$$
Here, we have used $|\Phi_M\psi>$ to denote the break down of the 
total wavefunction $\Phi_{M+1}$ into $\Phi_M$ and $\psi$. $x\equiv (r,\sigma)$, 
and r is the three dimensional Cartesian coordinate, and $\sigma$ is the spin index. 
Note that through the definition of Eq(8), $\Phi_{M+1}(x_1,x_2,...,x_M,x_{M+1})$
is antisymmetric, provided $\Phi_M$ is antisymmetric.  In general,
$\Phi_M(x_1,x_2,...,x_M)$ can be represented by a linear combination
of different electron configurations (i.e, single slater determinates from a single
particle  orthogonal basis set including $\psi$). 
Any slater determinate which consists of $\psi$ will be eliminated in Eq(8) by the
antisymmetry operation. Thus, it does not lose any generacity (in terms of variational
degree of freedom for $\Phi_{M+1}$) to 
exclude the single particle orbital $\psi$ from
the configurations of $\Phi_M(x_1,x_2,...,x_M)$. In another word, we can restrict
$\Phi_M(x_1,x_2,...,x_M)$ to be orthogonal to $\psi(x)$, i.e,

$$\int \Phi_M(x_1,x_2,...,x_M) \psi^*(x_1) d^3 x_1 =0.   \eqno(9) $$
This orthogonal condition is also called strong orthogonal condition \cite{Parr,book}.
Notice that, since we have not restrict any degree of freedom in $\Phi_{M+1}$ of Eq(8) 
by introducing Eq(9), Eq(9) should not be considered as an additional approximation. The
approximation, if any, has already been made in Eq(8). 
By using Eq(8), we have neglected the correlation between the single
particle state $\psi$ and the rest of the system $\Phi_M$. This is one approximate way to 
define an quasiparticle wavefunction $\psi(x_1)$ out of a many particle system. 
The approximation of Eq(8) has been called group function approximation \cite{Parr,book},
and has been used to study the effects (e.g, screening) of one group to another group. 
Thus, it is natural here to use it to describe the Coulomb screening effects of the background
system ($\Phi_M$) to the quasiparticle (and exciton) system ($\psi$).
Note that, Eq(8) goes beyond Hartree-Fock \cite{HF} approximation by retaining correlations within
$\Phi_M$. Since our focus 
is to study the dielectric screening of the Coulomb interaction, which can already be
described by Hartree Fock approximation, we expect Eq(8) to be
adequate to represent the main physics of interest here. 
Similar approximations of Eq(8)  has been used in the derivation
of Extended Koopmans' Theorem by Day, Smith, and Garrod \cite{Day}, by Morrell, 
Parr and Levy \cite{Morrell}, and by Kent, etal \cite{Kent}. In Eq(8), one gets 
the wavefunction $\Phi_{M+1}$ of the M+1 electron system by adding one electron $\psi$ 
to the M electron wavefunction $\Phi_M$. Similary, one can get the M electron system wavefunction
$\Phi_M$ by elminating $\psi(x_1)$ from  the M+1 electron wavefunction $\Phi_{M+1}$:

$$ \Phi_M(x_1,x_2,...,x_M) = \int \psi^*(x_{M+1}) \Phi_{M+1}(x_1,x_2,...,x_M,x_{M+1}) d^3 x_{M+1}.
  \eqno(10)$$
Using the orthogonal condition Eq(9), one can derive Eq(10) from Eq(8). Equation (10) 
is also used in the derivation of the Extended Koopmans' Theorem \cite{Day,Morrell,Kent}.

To appreciate the approximation embodied in Eq(8), we can take a look at the 
density matrix of the M+1 system. Under Eqs(8) and (9), $\rho_{M+1}(x,x')=
\psi(x)\psi^*(x')+\rho_M(x,x')$, here $\rho_{M+1}$ and $\rho_M$ are the density matrix of
the M+1 and M electron systems respectively. Besides, following the orthogonal condition
of Eq(9), $\int \rho_M(x,x') \psi(x') d^3 x'=0$. As a result, $\psi(x)$ is a natural 
orbital of $\rho_{M+1}$ with an occupation number of 1. As well known\cite{Kent},
the occupation
numbers of the natural orbitals are usually less than 1, thus the Eq(8) is definitely
an approximation. However, this approximation is  similar to the approximations
which use one single particle wavefunction to represent a quasiparticle in a many-body
system. These include: the Hartree-Fock theory \cite{HF}, empirical pseudopotential 
theory \cite{EPM}, the Kohn-Sham representation of the density functional 
theory \cite{KS} and in some extent the GW theory \cite{GW}.
To include the correlation between $\psi$ and $\Phi_M$ in Eq(8), one has
to add in Eq(8) other configurations $|\Phi'_M \psi'>$.  However, if the quasiparticle
is represented by a unique single particle wavefunction $\psi$, then $\psi'$ must be
$\psi$. As a result,
$|\Phi_M \psi>+|\Phi'_M \psi'>=|\Phi''_M \psi>$, where $\Phi''_M=\Phi_M+\Phi'_M$,
and Eq(8) remains unchanged.

The Hamiltonian for the M+1 particle system can be written as

$$H_{M+1}=-{1\over2} \sum_{i=1}^{M+1} \nabla^2_i + \sum_{i=1}^{M+1} 
V_{ion}(r_i)+ \sum_{i\ne j}^{M+1} {e^2\over |r_i-r_j|}. \eqno(11) $$
Using Eq(11) and Eq(8), we can evaluate $<\Phi_{M+1}|H_{M+1}|\Phi_{M+1}>$. 
After some algebras, it can be expressed as

$$<\Phi_{M+1}|H_{M+1}|\Phi_{M+1}>=E_{\psi}+<\Phi_M|H_M|\Phi_M> $$
$$~~~~~~+ \int \rho_M(r) V_{coul}^{\psi}(r) d^3 r
-\int V_{ex}^{\psi}(x,x^{\prime})\bar\rho_M(x,x^{\prime}) d^3x d^3x^{\prime},   \eqno(12) $$
where $H_M$ is just the $H_{M+1}$ of Eq(11), but change the $M+1$ to M. 
$E_{\psi}$ is defined as

$$E_{\psi}=\int \psi(x) \left [ -{1\over2} \nabla^2+V_{ion}(r) \right ] \psi^*(x) d^3x
. \eqno(13)$$
$V_{coul}^{\psi}(r)$ is the Coulomb potential due to $\psi$,

 $$V_{coul}^{\psi}(r)=\int {e^2\over |r-r^{\prime}|} |\psi(x^{\prime})|^2 d^3 x^{\prime}, \eqno(14)$$  
$\rho_M(r)$ is the total charge density of $\Phi_M$,

 $$\rho_M(r)= M \sum_{\sigma}
 \int |\Phi_M((r,\sigma),x_2,...x_M)|^2 d^3x_2...d^3x_M,  \eqno(15) $$
$V_{ex}^{\psi}(x,x^{\prime})$ is the exchange nonlocal potential due to 
$\psi$,

$$ V_{ex}^{\psi}(x,x^{\prime}) = {e^2\over |r-r^{\prime}|}
  \psi^*(x) \psi(x^{\prime}), \eqno(16)$$        
and $\bar \rho_M(x,x^{\prime})$ is the density matrix of $\Phi_M$,

$$\bar \rho_M(x,x^{\prime}) = M \int \Phi_M(x,x_2,..,x_M)
 \Phi_M^*(x^{\prime},x_2,..,x_M) d^3x_2...d^3x_M.  \eqno(17)  $$
  Note that $\rho_M(r)=\sum_{\sigma} \bar\rho_M(r\sigma,r\sigma)$.

Equation (12) has been derived before as in the group function theory \cite{book}.
Under the assumption of the wavefunction $\Phi_{M+1}$ of Eq(8), one can variationally minimize
the total energy of Eq(12) with regard to $\Phi_M$ and $\psi$ under the constraint of
Eq(9). 
Doing so, both $\Phi_M$ and $\psi$ can be solved selfconsistently from Eq(12). However, 
symmetries might need to be used to keep the $\psi$ to be the desired single particle wavefunctions
(i.e, the top of valence state and bottom of conduction state). Another approach is to calculate
$\psi$ from single particle Hamiltonians, e.g, quasi-particle GW method \cite{GW}, or empirical 
pseudopotential method \cite{EPM}.
Later in Section IV, we will provide a new single particle Hamiltonian
$H_s$ [Eq(40)] to calculate $\psi$.
For a given $\psi$, we can minimize 
Eq(12) with regard to $\Phi_M$, thus get a many particle equation for $\Phi_M$ under the
perturbation of Coulomb potential
$V_{coul}^{\psi}(r)$ and the exchange potential $V_{ex}^{\psi}(x,x^{\prime})$.

\vskip 0.5in
\centerline{\bf III. Coulomb energy associated with the ionization 
energy and electron affinity}
\vskip 0.2in

 The exciton energy $E_{exc}$ can be directly defined as 
 $E_N^*-E_N$. Here, $E_N$ is the N electron ground state energy and
 $E_N^*$ is the total energy of the N electron
 system which contains one exciton (i.e, an conduction band single particle state $\psi_c$ 
 is occupied and an valence band single particle state 
 $\psi_v$ is unoccupied).
Then the Coulomb energy which connects $E^{qp}_g$ ($\equiv 
E_{N+1}+E_{N-1}-2E_N$) with $E_{exc}$ is

 $$ E_{coul}\equiv E_{exc}-E^{qp}_g = (E_N^*-E_{N-1})-(E_{N+1}-E_N),     \eqno(18) $$ 
 where, $E_{N-1}$ and $E_{N+1}$ are the ground state
 energies of N-1 and N+1 electron systems. Note that in $E_N^*$, $E_N$,
 $E_{N-1}$ and $E_{N+1}$ systems, the ionic potential $V_{ion}(r)$ in Eq(11) is 
 the same. The difference is the electron occupations. Using the approach 
 outlined in the previous section, we will approximate the wavefunctions of
 $E_N^*$, $E_N$, $E_{N-1}$ and $E_{N+1}$ systems as 
 $|\Phi^c_{N-1}\psi_c>$, $|\Phi^0_N>$, $|\Phi^0_{N-1}>$ and $|\Phi^c_N \psi_c>$
 respectively.  Using the orthogonal conditions of Eq(9), and applying it equally
 to all the systems in Eq(18), we have $\Phi^c_{N-1}$, $\Phi^0_N$, $\Phi^0_{N-1}$
 and $\Phi^c_N$ all orthogonal to $\psi_c$ [Eq(9), replacing $\psi$ by $\psi_c$,
 and $\Phi_M$ by these four wavefunctions]. In addition, $\Phi^c_{N-1}$ and
 $\Phi^0_{N-1}$ should be orthogonal to $\psi_v$ [Eq(9), replacing $\psi$ by $\psi_v$], 
 so that a hole exists in these systems. Notice that, we have used the superscription
 ``0'' to indicate that the corresponding wavefunction $\Phi_M^0$ is a minimum energy 
 variational solution of $<\Phi_M^0|H_M|\Phi_M^0>$. Similarly, the superscription 
 ``c'' indicates that the corresponding wavefunction $\Phi_M^c$ is the variational 
 solution of Eq(12) under the perturbation of $\psi_c$ 
 [through $V_{coul}^{\psi_c}$ and $V_{ex}^{\psi_c}$].
Notice that, only in the strong confinement quantum dot size region \cite{Yoffe}, we 
can write $\Phi_N^*$ as $|\Phi^c_{N-1}\psi_c>$ through Eq(8). In this size region, 
the electron and the hole are not correlated. 
As a result, we don't have to write $\Phi^*_N$ as a summation of configurations
$|\Phi^c_{N-1}\psi_c>$ with different electron ($\psi_c$) and hole ($\psi_v$) wavefunctions. 
After the definitions of the wavefunctions, we can replace $E_N^*$, $E_N$, $E_{N-1}$ and $E_{N+1}$
by $<\Phi^c_{N-1}\psi_c| H_N |\Phi^c_{N-1}\psi_c>$, $<\Phi^0_N|H_N|\Phi^0_N>$,
$<\Phi^0_{N-1}|H_{N-1}|\Phi^0_{N-1}>$ and $<\Phi^c_N \psi_c|H_{N+1}|\Phi^c_N \psi_c>$,
respectively. Using Eq(12), we have:

$$E_N^*-E_{N-1}= E_{\psi_c}+<\Phi^c_{N-1}|H_{N-1}|\Phi^c_{N-1}>
  -<\Phi_{N-1}^0|H_{N-1}|\Phi_{N-1}^0>      $$
$$+ \int \rho^c_{N-1}(r) V^{\psi_c}_{coul}(r) d^3 r -\int \bar \rho^c_{N-1}(x,x')
V_{ex}^{\psi_c}(x,x^{\prime}) d^3 x d^3 x^{\prime},   \eqno(19)  $$
and
$$E_{N+1}-E_N=  E_{\psi_c}+<\Phi^c_N|H_N|\Phi^c_N>
  -<\Phi_N^0|H_N|\Phi_N^0>      $$
$$+ \int \rho^c_N(r) V^{\psi_c}_{coul}(r) d^3 r -\int \bar \rho^c_N(x,x')
V_{ex}^{\psi_c}(x,x^{\prime}) d^3 x d^3 x^{\prime}.   \eqno(20)  $$

The changes from $\Phi_{N-1}^0$, $\Phi_N^0$ to $\Phi^c_{N-1}$, $\Phi^c_N$ can be
described in first order by perturbation theory, under the external perturbation
potentials $V^{\psi_c}_{coul}(r)$ and $V_{ex}^{\psi_c}$ in Eq(12).
For a N-1 electron system, the energy changes from
$<\Phi^0_{N-1}|H_{N-1}|\Phi^0_{N-1}>$ to $<\Phi_{N-1}^c|H_{N-1}|\Phi_{N-1}^c>$ 
in response to these perturbative  potentials. The amplitude of this energy
change is in the same order of the Coulomb
interaction of the perturbation charge ($\psi_c^2$)
with the response charge of the system (which is also in the order of $\psi_c^2$).
In other words, 
$\Delta E_{N-1}\equiv <\Phi^c_{N-1}|H_{N-1}|\Phi^c_{N-1}>-<\Phi_{N-1}^0|H_{N-1}|\Phi_{N-1}^0>$  scales
as 1, instead of N. This is also evident from Eq(19). Note that $E_N^*-E_{N-1}$ 
scales as 1. At the right hand side of Eq(19), combining the $E_{\psi_c}$ term defined in Eq(13) with the
$\int \rho^c_{N-1}(r) V^{\psi_c}_{coul}(r) d^3 r -\int \bar \rho^c_{N-1}(x,x')
V_{ex}^{\psi_c}(x,x^{\prime}) d^3 x d^3 x^{\prime}$ term, one gets an energy similar
to that of the Hartree Fock single particle eigen energy. Thus, this combined term also
scales as 1. As a result, the $\Delta E_{N-1}$ scales as 1. 
The same is true for 
$\Delta E_N \equiv <\Phi^c_N|H_N|\Phi^c_N>-<\Phi_N^0|H_N|\Phi_N^0>$.
The difference between $\Delta E_{N-1}$ and $\Delta E_N$ is the difference between 
N-1 and N electron system, which should scale as $\Delta E_N/N \propto 1/N$. Thus, this difference 
is negligible comparing to the Coulomb 
energy $E_{coul}$ which we are calculating. As a result, subtract Eq(20) from Eq(19), we have 

$$E_{coul}=\int [\rho^c_{N-1}(r)-\rho^c_N(r)] V_{coul}^{\psi_c}(r) d^3 r $$
$$~~~~~~~~ -\int \left [\bar \rho^c_{N-1}(x,x^{\prime})-\rho^c_N(x,x^{\prime})
   \right ] V_{ex}^{\psi_c}(x,x^{\prime}) d^3 x d^3 x^{\prime}. \eqno(21)$$
Further more, following the same perturbation argument, we have
  $\rho^c_N(r)-\rho_N^0(r)\simeq \rho^c_{N-1}(r)-\rho_{N-1}^0(r)$ and
  ${\bar \rho}^c_N(x,x^{\prime})-{\bar \rho}_N^0(x,x^{\prime})\simeq
  {\bar \rho}^c_{N-1}(x,x^{\prime})-{\bar \rho}_{N-1}^0(x,x^{\prime})$. 
Again, the approximation has an relative error $\propto 1/N$, thus negligible. 
Then, we can change Eq(21) to 

$$E_{coul}=\int [\rho_{N-1}^0(r)-\rho_N^0(r)] V_{coul}^{\psi_c}(r) d^3 r
 -\int \left [{\bar \rho}_{N-1}^0(x,x^{\prime})-{\bar \rho}_N^0(x,x^{\prime})
 \right ] V_{ex}^{\psi_c}(x,x^{\prime}) d^3 x d^3 x^{\prime}. \eqno(22)$$
This is our central result. 
Notice that, $\rho_{N-1}^0(r)$ is the total electron charge density of a N-1 electron system
(containing 
one hole), and $\rho_N^0(r)$ is the total electron charge density of a N electron neutral system. Thus,
$\rho^{src}_v(r)\equiv \rho_{N-1}^0(r)-\rho_N^0(r)$ is the screened hole charge
(the bare hole charge plus the response screening charge) in the quantum dot. 
Note that

$$ \int \rho^{src}_v(r) d^3r = -1.    \eqno(23) $$
Thus, if we ignore the exchange interaction in Eq(22) (which is usually much smaller than the
Hartree interaction), and express $V_{coul}^{\psi_c}(r)$ using Eq(14),  we have

$$ E_{exc} = E^{qp}_g + \int {\rho^{src}_v(r_1) \rho_c(r_2) \over |r_1-r_2| } d^3r_1 d^3r_2 \eqno(24) $$ 
where, $\rho_c(r)=\sum_{\sigma} |\psi_c(r\sigma)|^2$.
In the case where phenomenological dielectric constant $\epsilon$ can be used to describe $\rho^{src}_v(r_1)$,
we found that Eq(24) is in agreement with Eq(6), not Eq(7). More, explicitly, in that case
we have

$$\rho_v^{scr}({\bf r})= -{1\over \epsilon} \rho_v(r)  ~~~~ for ~~ r<R$$
$$ ~~~~~~~~~~~~~~~~~~~=-(1-{1\over \epsilon})\delta(r-R)/4\pi R^2  ~~~ for ~~
r \sim R $$
$$ ~~~~~~~~~~~~~~~~~~=0 ~~~~~~~~~~~~~~~~~ for ~~~~ r > R, \eqno(25) $$
where $R$ is the quantum dot radius and $\rho_v(r)=\sum_{\sigma}|\psi_v(r\sigma)|^2$.
Using the effective mass charge density $\rho_v(r)=\rho_c(r)=sin^2(\pi r/R)/2\pi R r^2$, we yield 
Eq(6) from Eq(24).

  In the empirical electrostatic derivation of Eqs(1)-(2), it is not so clear whether the 
  empirical single particle eigenvalues $\epsilon_e$, $\epsilon_h$ [in Eqs(1)-(2)] 
  already include the effects of electrostatic energies. For example, when one electron
  is removed from the quantum dot, how to relate the single electron eigenvalue and
  classical electrostatic energy to the change of the total energy in the system is 
  problematic \cite{Brus1,Brus2}. Here, deriving Eq(24) from the many-body wavefunctions, we do not have
  all these conceptional difficulties. That is the biggest advantage of the current 
  derivation comparing to the classical empirical derivations \cite{Brus1,Brus2}.
   Notice that, the dielectric constant $\epsilon$ is not derived here, unlike in the case of
bulk exciton screening\cite{Sham}.
This, of course,  doesn't mean that we have no screening in our formalism.
The dielectric screening of the Coulomb interaction is enclosed
in $\rho_v^{src}$ of Eq(24) in our formula. 
We just didn't derive the detail screening function in our formula (i.e, how
to calculate $\rho_v^{src}$ from $\rho_v$ by a detail $\epsilon$ model). That is
not the focus of our current study.
In the cases of Eqs(6) and (7), the simple response model of Eq(25) is assumed. It is 
in this context, $\epsilon$ is used here, and Eq(6) is found to be correct.

\vskip 0.5in
\centerline{\bf IV. Coulomb energy associated with the single particle
eigenvalues}
\vskip 0.2in

 Had proved Eq(6), now we like to test Eq(5). What important here is to 
 define an effective single particle Hamiltonian, for which  $\epsilon_e$ and 
 $\epsilon_v$ are its conduction band minimum and valence band maximum eigen
 energies.
Let's start from the definition of the exciton energy:
$E_{exc}=E_N^*-E_N$. Now, we will rewrite the N particle ground
state wavefunction $\Phi_N^0$ of $E_N$ as $|\Phi^v_{N-1}\psi_v>$, with 
$\Phi^v_{N-1}$ being orthogonal to both $\psi_v$ and $\psi_c$. Using
Eq(12), $E_N$ can be written as:

 $$E_N-E_{\psi_v}=<\Phi^v_{N-1}|H_{N-1}|\Phi^v_{N-1}> $$
 $$~~~~~ +\int \rho_{N-1}^v(r) V_{coul}^{\psi_v}(r) d^3 r
   -\int \bar \rho_{N-1}^v(x,x') V_{ex}^{\psi_v}(x,x') d^3x d^3x'.  \eqno(26)$$
Here, $E_{\psi_v}$ is evaluated from Eq(13). 
Similarly, like before, the exciton wavefunction can be 
expressed as $|\Phi^c_{N-1}\psi_c>$, with $\Phi^c_{N-1}$ being orthogonal to $\psi_v$
and $\psi_c$. Then, $E_N^*$ can be written as:

$$E_N^*-E_{\psi_c}=<\Phi^c_{N-1}|H_{N-1}|\Phi^c_{N-1}>$$
$$~~~~~+\int \rho^c_{N-1}(r) V_{coul}^{\psi_c}(r) d^3 r
  -\int \bar \rho^c_{N-1}(x,x') V_{ex}^{\psi_c}(x,x') d^3x d^3x'.   \eqno(27) $$
Comparing Eq(27) with Eq(26), we find that we can obtain $\Phi^c_{N-1}$ 
from $\Phi_{N-1}^v$ by applying potentials
 $V_{coul}^{\psi_c}-V_{coul}^{\psi_v}$ and $V_{ex}^{\psi_c}-V_{ex}^{\psi_v}$
 into Eq(26). More specifically we can define

$$E(\beta)\equiv <\Phi^{\beta}_{N-1}|H_{eff}^{\beta}|\Phi^{\beta}_{N-1}>= $$
$$<\Phi^{\beta}_{N-1}|H_{N-1}+ 
[(1-\beta) V_{coul}^{\psi_v}+\beta V_{coul}^{\psi_c}]
 -[(1-\beta) V_{ex}^{\psi_v}+\beta V_{ex}^{\psi_c}]|\Phi^{\beta}_{N-1}>. \eqno(28)$$
In Eq(28), $\Phi^{\beta}_{N-1}$ is defined as the minimum energy variational 
solution of the Hamiltonian $H_{eff}^{\beta}$, while subjected to orthogonal
condition Eq(9) to $\psi_c$ and $\psi_v$.
Using Eq(26) and Eq(27), we have $E(\beta=0)=E_N-E_{\psi_v}$ and $E(\beta=1)=E_N^*-E_{\psi_c}$.
Now, using the ``adiabatic integration technique'', we have

$$E(\beta=1)=E(\beta=0)+\int_0^1 {\partial E(\beta)\over \partial \beta}
 d \beta .    \eqno(29)$$
Note, in Eq(28), $\Phi^{\beta}_{N-1}$ is the variational solution of 
$H_{eff}^{\beta}$, as a result,

$$<{\partial \Phi^{\beta}_{N-1}\over \partial \beta}|H_{eff}^{\beta}|\Phi^{\beta}_{N-1}>
=<\Phi^{\beta}_{N-1}|H_{eff}^{\beta}
|{\partial \Phi^{\beta}_{N-1}\over \partial \beta}> =0.    \eqno(30)$$
Then, we have
$$E(\beta=1)=E(\beta=0)+\int_0^1 <\Phi^{\beta}_{N-1}|{\partial H_{eff}^{\beta}
\over \partial \beta}|\Phi^{\beta}_{N-1}> d\beta    $$
$$~~~~~=E(\beta=0)+\int_0^1  \Big\{ \int
\rho_{N-1}^{\beta}(r) \left [V_{coul}^{\psi_c}(r)-\int V_{coul}^{\psi_v}(r) \right ] d^3 r   $$
$$-\int \bar \rho_{N-1}^{\beta}(x,x') \left [
V_{ex}^{\psi_c}(x,x')-\int V_{ex}^{\psi_v}(x,x') \right ] d^3x d^3x'
\Big\} d\beta .   \eqno(31)$$
$\rho_{N-1}^{\beta}$ and $\bar \rho_{N-1}^{\beta}$ are the density
and density matrix of the N-1 electron system under the perturbation of 
$\beta [V_{coul}^{\psi_c}- V_{coul}^{\psi_v}]$ and
$\beta [V_{ex}^{\psi_c}- V_{ex}^{\psi_v}]$ (the nonperturbated values are
$\rho_{N-1}^v$ and $\bar \rho_{N-1}^v$). These perturbative potentials are due to 
charge density $\beta [|\psi_c(x)|^2-|\psi_v(x)|^2]$ and density matrix
$\beta [ \psi_c(x)\psi_c^*(x')-\psi_v(x)\psi_v^*(x')]$. Here, we will only
consider cases where a phenomenological macroscopic dielectric constant
$\epsilon$ can be used to describe the response of the N-1 electron
system to these perturbative charges. These perturbative charges 
can be considered as external to the N-1 electron system. 
Thus, under this phenomenological
description, the N-1 electron charges can be expressed as

$$\rho_{N-1}^{\beta}(r)=\rho_{N-1}^v(r)+\beta ({1\over\epsilon}-1)
 \sum_{\sigma}[|\psi_c(r,\sigma)|^2-|\psi_v(r,\sigma)|^2]
 \eqno(32) $$
$$\bar \rho_{N-1}^{\beta}(x,x')=\bar \rho_{N-1}^v(x,x')+\beta ({1\over\epsilon}-1)
 [\psi_c(x)\psi_c^*(x')-\psi_v(x)\psi_v^*(x')]
 \eqno(33) $$
Note that, $\rho_{N-1}^{\beta}(r)=\sum_{\sigma}\rho_{N-1}^{\beta}
[(r,\sigma),(r,\sigma)]$, thus the dielectric constant $\epsilon$ used to describe
the response of $\rho_{N-1}^{\beta}(r)$ [Eq(32)] 
and $\bar \rho_{N-1}^{\beta}(x,x')$ [Eq(33)] must be the same. 
Also notice that, the dielectric constant $\epsilon$ used in Eqs(32),(33) is consistent
with the conventional definition of $\epsilon$, and there is no surface response charge
as in Eq(25) because the net charge of $\sum_{\sigma}[|\psi_c(r,\sigma)|^2-|\psi_v(r,\sigma)|^2]$ is zero.
Now, substitute Eqs(32)-(33) into Eq(31), carry out the $\beta$ integration, 
we have:

$$E_N^*-E_{\psi_c}=E_N-E_{\psi_v}  $$
$$ +\int \Big\{ \rho_{N-1}^v(r)+{1\over2}({1\over\epsilon}-1)
  \sum_{\sigma}[|\psi_c(r,\sigma)|^2-|\psi_v(r,\sigma)|^2]
\Big\} [V_{coul}^{\psi_c}(r)-V_{coul}^{\psi_v}(r)] d^3r  $$
$$+\int \Big\{ \bar \rho_{N-1}^v(x,x')+{1\over2}({1\over\epsilon}-1)
 [\psi_c(x)\psi_c^*(x')-\psi_v(x)\psi_v^*(x')] \Big\}
 [V_{ex}^{\psi_c}(x,x')-V_{ex}^{\psi_v}(x,x')] d^3x d^3x'.    \eqno(34) $$
Note that $\rho^0_N(r)=\rho^v_{N-1}(r)+\sum_{\sigma} |\psi_v(r,\sigma)|^2$
and $\bar \rho^0_N(x,x')=\bar \rho^v_{N-1}(x,x')+\psi_v(x)\psi_v^*(x')$, 
where $\rho^0_N(r)$ and $\bar \rho^0_N(x,x')$ are the neutral system N electron ground
state charge density and density matrix. Using this relations, and 
$E_{exc}=E_N^*-E_N$, we can derive from Eq(34) that:

$$E_{exc}=E_{\psi_c}-E_{\psi_v}+
 \int \rho_N^0(r) [V^{\psi_c}_{coul}(r)-V^{\psi_v}_{coul}(r)] d^3 r 
-\int \bar \rho_N^0(x,x') 
 [V^{\psi_c}_{ex}(x,x')-V^{\psi_v}_{ex}(x,x')] d^3 x d^3 x'     $$
$$-{1\over\epsilon} \int {|\psi_v(x)|^2 |\psi_c(x')|^2\over |r-r'|} d^3x d^3x' 
+{1\over\epsilon} \int {\psi_c(x)\psi_v(x) \psi_c^*(x')\psi_v^*(x')
\over |r-r'|} d^3x d^3x'  \eqno(35) $$
Similar to Eqs(14) and (16), 
we can now define Coulomb and exchange potentials due to the N electron ground
state charge density $\rho_N^0(r)$ and density matrix $\bar \rho_N^0(x,x')$:

 $$V_{coul}^N(r)=\int {1\over |r-r'|} \rho_N^0(r') d^3 r'   \eqno(36) $$
 $$V_{ex}^N(x,x')={1\over |r-r'|} \bar\rho_N^0(x,x')   \eqno(37)$$
Then, we can defined a effective single particle Hamiltonian $H_s$:

 $$\epsilon_{\psi}\equiv <\psi|H_s|\psi>= $$
$$~~~~~~~~ \int \psi^*(x) \Big\{ \delta(x-x')[-{1\over2} \nabla^2+V_{ion}(r)+
 V_{coul}^N(r)]-V_{ex}^N(x,x') \Big\} \psi(x') d^3x d^3x'.  \eqno(38)$$
Using Eq(38), and Eqs(14),(16), the first part of Eq(35) can be 
simplified, and it leads to 

$$E_{exc}=\epsilon_{\psi_c}-\epsilon_{\psi_v} -
{1\over\epsilon} \int {|\psi_v(x)|^2 |\psi_c(x')|^2\over |r-r'|} d^3x d^3x' $$
$$+{1\over\epsilon} \int {\psi_c(x)\psi_v(x) \psi_c^*(x')\psi_v^*(x')
\over |r-r'|} d^3x d^3x', \eqno(39) $$ 
where $\epsilon_{\psi_c}$ and $\epsilon_{\psi_v}$ are the eigenvalues of the
bottom of conduction band and top of valence band of the effective single
particle Hamiltonian $H_s$ in Eq(38). 
This concludes our major result for this section. 

Comparing Eq(39) to Eq(5), we notice that:
(1) The bulk band gap $E^{bulk}_g$ in Eq(5) has been absorbed in the definition of the
single particle eigen values $\epsilon_{\psi_c}$ and $\epsilon_{\psi_v}$. (2) The last
term in Eq(5), which represents the classical instantaneous dielectric energy, does
not exist in the current result. It is thus quite plausible that this term should 
not exist in Eq(4) either, which affects the electron affinity [Eq(1)], ionization
energy [Eq(2)] and quasi-particle energy [Eq(3)].

Equation (39) confirms the conventional way to calculate 
the exciton energy via the single particle eigen values, e.g, as in the empirical
pseudopotential approach \cite{EPM}. Both the Hartree Coulomb interaction and the exchange interaction 
exist in Eq(39). Interestingly, following the assumption of Eqs(32)-(33),
the exchange interaction is screened in Eq(39) just like the Coulomb interaction. This might
shed some light on the long standing controversy about whether the exchange interaction 
should be screened \cite{Abe,Sham,exchange} [notice also the second term in Eq(22)]. 
Also important here is that we have provided a definition of an effective single
particle Hamiltonian $H_s$ in Eq(38) using the N electron charge density and 
density matrix. Equation (38) is like the variational equation obtained from Eq(12) for a given
$\Phi_M$. The difference is that, here, the same $\Phi_N$ is used for both 
$\psi_v$ and $\psi_c$.
This $H_s$ can be compared with the conventional single particle Hamiltonians
e.g, EPM \cite{EPM}, local density approximation (LDA) \cite{KS}, quasi-particle GW calculation
\cite{GW} and Hartree-Fock equation \cite{HF}. 
$H_s$ is almost the same as the Hartree-Fock equation, but that the uncorrelated
HF exchange potential $V_{ex,HF}^{N}(x,x')={1\over |r-r'|} \sum_{i\in occ} \psi_i(x) \psi_i^*(x')$
has been changed to the correlated exchange potential $V_{ex}^N(x,x')$ of Eq(37).
According to Eq(39), the eigenvalues of $H_s$ should
provide the  band structure (at least the band gap) of a bulk system. This 
remains to be tested.

Equations (6),(24) and Eq(39) represents two different approachs to calculate the exciton energy in a
nanostructure. In the paper of Ogut etal \cite{Ogut}, the first approach is taken. Unfortunately, Eq(7) instead of
Eq(6) was used. If Eq(6) were used, the result of Ogut etal would be very close (within $\sim 0.1$ eV) to the results
obtained via EPM calculations \cite{EPM} and Eq(39) 
(a constant 0.68 eV Si LDA band gap correction needs to be added to $E^{qp}_{g,LDA}(R)$ in order to 
get this good agreement).

\vskip 0.5in
\centerline{\bf V. Numerical test of the formulae}
\vskip 0.2in

  Using local density approximation method, 
  Ogut, etal \cite{Ogut} have calculated the $E_{N+1}+E_{N-1}-2E_N$ and the single particle
  Kohn-Sham eigenvalues. If we approximate the Kohn-Sham eigenvalue as the single
  particle energies $\epsilon_{\psi_c}$($\equiv\epsilon_c$) and $\epsilon_{\psi_v}$($\equiv\epsilon_v$)
  given by Eq(38), then we have
  an good case to test our equations (24) and (39).  According to Eqs(24) and (39), 
  we have (ignoring the exchange interactions)

$$ \Sigma \equiv (E_{N+1}+E_{N-1}-2 E_N)-(\epsilon_c-\epsilon_v)  
=  (1-{1\over\epsilon(R)}) {1\over R}  \eqno(40) $$
We have used the same symbol $\Sigma$ in Eq(40) as in Ref.\cite{Ogut}, although we do not 
mean that it is the GW ``selfenergy'' as claimed in Ref.\cite{Ogut}. According to Eq(40), 
$\Sigma$ is simply an electrostatic energy between the electron charge and its induced
surface charge. Although there is a well know LDA error for the band gap energy, 
this error exists in both $(E_{N+1}+E_{N-1}-2 E_N)$ and $(\epsilon_c-\epsilon_v)$, thus
should be cancelled. As a result, Eq(40) should still be valid for LDA calculations. 

 In Fig.1, the quantity $(E_{N+1}+E_{N-1}-2 E_N)-(\epsilon_c-\epsilon_v)$ is plotted as a function
 of 1/R. This quantity is compared with $(1-{1\over\epsilon(R)}){1\over R}$. The
 agreement is quite good. Here, $\epsilon(R)$ is a function of the quantum dot
 radius R. This function is calculated in Ref.\cite{PRLW}, and can be
 expressed as $\epsilon(R)=1+(11.4-1)/(1+(\alpha/R)^l)$. 
 Here, we have used
 $\epsilon(R)$ which corresponds to the total polarizibility of the 
 quantum dot ($\epsilon_s$ in Ref.\cite{PRLW}), i.e, $\alpha=4.25\AA$ and
 $l=1.25$.  Of course, the agreement of 
 Eq(40) depends on how good is the phenomenological macroscopic description of the quantum
 dot screening [Eq(25)]. There is no reason to believe that Eq(25) is exact for a
 small quantum dot. After considered all these uncertainties, the
 agreement in Fig.1 is quite good.

 Notice that, the LDA Kohn-Sham single particle band gap $\epsilon_c-\epsilon_v$ is almost
 the same (within 0.1 eV) as the single particle band gap calculated from empirical pseudopotential 
 after a 0.68 eV  band gap
 correction is added to LDA result. This can be confirmed by taken the data
 from Ref.\cite{Ogut} and Ref.\cite{EPM}. It has also been confirmed separately by Delley etal
 in Ref.\cite{Delley}. 
 Then, using the good agreement between LDA $(E_{N+1}+E_{N-1}-2E_N)-(\epsilon_c-\epsilon_v)$
 and $(1-1/\epsilon(R))/R$ as shown in Fig.1, we know that the exciton energy $E_{exc}$ calculated
 from LDA quasiparticle energy through Eq(6) [plus 0.68 eV correction], should be the same as $E_{exc}$
 calculated from empirical
 pseudopotential single particle eigen values through Eq(40) \cite{EPM}.

\vskip 0.5in
\centerline{\bf VI. Conclusions}
\vskip 0.2in

   One approximation [Eq(8)] to the many-body wavefunction is presented. This approximation 
allows us to define a single particle wavefunction in a many-body system. This 
approximation is more accurate than the Hartree-Fock Slater determinate for the many-body 
wavefunction. Under this approximation, 
 we have derived the Coulomb energy
needed to relate the quasiparticle energy $E_g^{qp}$ to the exciton energy $E_{exc}$
(= the optical transition energy)  [Eqs(22),(24)]. In the limit
where the phenomenological description of the dielectric screening is valid, we found that the
correct formula for this Coulomb energy is Eq(6), not Eq(7). 
We also derived the Coulomb energy which relates the single particle eigen values
with the $E_{exc}$ [Eq(39)]. We found that the classical instantaneous dielectric energy
in the last term of Eq(5) does not exist in our currently derived formula [Eq(39)]. Under the 
assumption of Eqs(32)-(33), we found that the exchange interaction in Eq(39) is screened
as for the Hartree interaction.  Using Eq(24) and Eq(39) respectively, we found that the optical 
transition energy obtained from the LDA $E_g^{qp}$ 
(plus the LDA band gap correction), 
is almost the same as the result obtained from 
the EPM single particle eigen values.
Thus, the conclusion made in Ref.\cite{Ogut}, that EPM misses the change of selfenergy with size R, is incorrect.
Finally, we presented an effective single particle Hamiltonian $H_s$ [Eq(38)], which under the
assumption of Eq(8), provides the band structure of a bulk system. Further testing of
Eq(38) is needed.

  I thank my colleagues A. Zunger, A. William, A. Franceschetti and R. Needs for useful discussions. 
  This work was supported by the Office of Energy Research, Material Science
  Division, U.S. Department of Energy, under Grant No. DE-AC02-83CH10093.

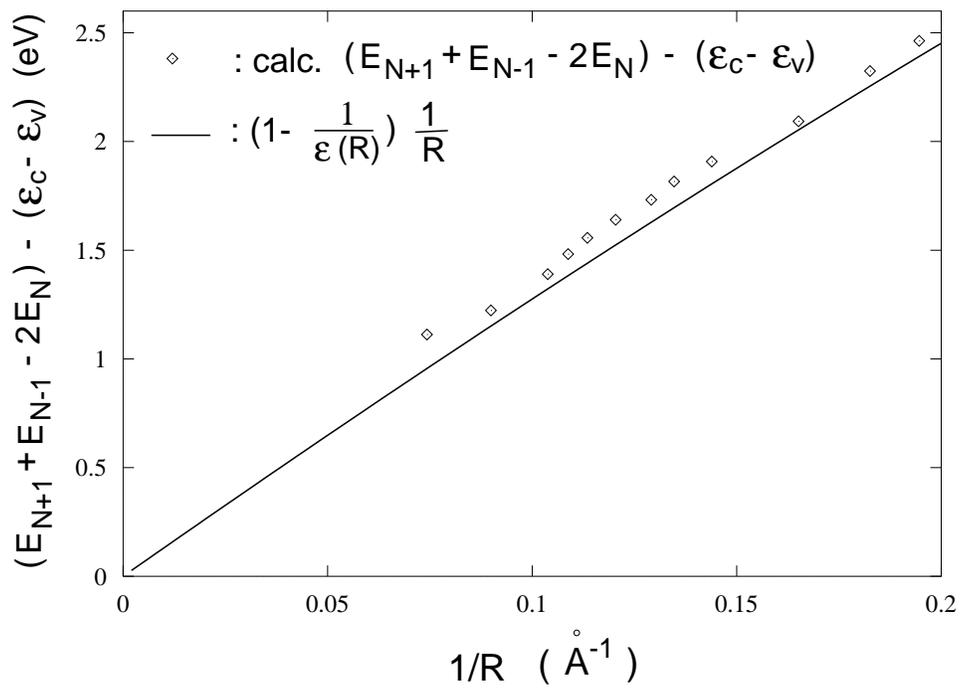
\begin {figure}
\caption {The  LDA calculated $(E_{N+1}+E_{N-1}-2 E_N) -(\epsilon_c-\epsilon_v)$
compared with the surface polarization Coulomb interaction $(1-1/\epsilon(R))/R$.}
\end {figure}

\end{document}